\documentclass{mn2e}

\usepackage{psfig}

\newif\ifAMStwofonts

\def\til{$\sim$}

\def\ha{H$\alpha$\space}
\def\hb{H$\beta$\space}

\def\he{He\thinspace{\sc i}\thinspace{$\lambda$}4471\space}
\def\heii{He\thinspace{\sc ii}\thinspace{$\lambda$}4686\space}
\def\civ{C\thinspace{\sc iv}\thinspace{$\lambda$}1550\space}
\def\th{\thinspace}

\title{Mass Outflow from the X-ray Binary X2127+119 in M15}
\author[L. van Zyl, T. Naylor, P.A. Charles, Z. Ioannou]
       {L. van Zyl$^{1,2}$\thanks{Present address: Astrophysics Group, School of Chemistry and Physics, Keele University, Staffordshire, ST5 5BG, United Kingdom. E-mail: lvz@astro.keele.ac.uk}, T. Naylor$^{2,4}$, P.A. Charles$^{1,3}$, Z. Ioannou$^{2,5}$ \\
       $^1$Department of Astrophysics, Oxford University, Keble Road, 
           Oxford OX1 3RH\\
       $^2$Astrophysics Group, School of Chemistry and Physics, Keele University, Staffordshire, ST5 5BG\\
       $^3$Department of Physics \& Astronomy, University of
           Southampton, Southampton SO17 1BJ\\
       $^4$School of Physics, University of Exeter, Exeter EX4 4QL\\
       $^5$Department of Astronomy, University of Texas at Austin, Austin, TX
           78712, USA
       }
\date{Accepted 
      Received }

\pagerange{\pageref{firstpage}--\pageref{lastpage}}
\pubyear{2002}

\begin{document}

\maketitle

\label{firstpage}

\begin{abstract}

The bright X-ray binary X2127+119 in the core of the globular cluster M15 has long been thought to be in an unusual evolutionary state, in which the binary is embedded in a common envelope. Support for this idea comes from X2127+119's absorption lines, which are blue shifted at all orbital phases, indicating the existence of outflows from the system. A common-envelope scenario implies that the absorption lines should exhibit maximum blue shift near mid-eclipse (binary phase 0.0). We have re-analysed INT spectra of X2127+119 obtained in 1986, 1987 and 1988 using the latest orbital ephemeris (substantially different from that used in the original analysis), and find that the orbital phase at which the absorption lines show a maximum blue shift is not 0.0, but rather 0.25 -- 0.3. These results indicate that a common-envelope scenario for X2127+119 may not work. In addition, from spectrograms of the \heii line, we report the first tentative detection of X2127+119's companion star.
\end{abstract}

\begin{keywords}
Accretion -- Stars: X-ray binaries -- Stars: individual: X2127+119, AC211
\end{keywords}

\section{The X-ray binary in M15}

The X-ray binary X2127+119 (AC211) in the core of the globular cluster M15 is optically one of the brightest and most well-studied of the low-mass X-ray binaries (LMXBs), and yet it remains a highly enigmatic system. Its optical and X-ray light-curves, along with its very low $L_{\rm X}$/$L_{\rm opt}$, make it a classic accretion disc corona (ADC) source (Fabian, Guilbert \& Callanan 1987; Naylor et al. 1988), in which the system is seen almost edge-on and the compact object and hot, luminous inner disc are obscured by the accretion disc rim. The X-ray flux we observe comes entirely from photon-scattering by a large corona above the disc, and is only a small fraction of the source's intrinsic X-ray flux.

The X-ray source X2127+119 has recently been discovered by {\it Chandra} to be two separate sources (White \& Angelini 2001). The second X-ray binary is 2.7 arcsec from AC211, is 4 magnitudes fainter in the U-band, but its {\it Chandra} count-rate is 2.5 times higher. The discovery of a second LMXB in the core of M15 explains one of most puzzling aspects of AC211: luminous X-ray bursts showing expansion of the neutron star photosphere have been observed in X2127+119 (Dotani et al. 1990, van Paradijs et al. 1990, Smale 2001), which has been difficult to reconcile with AC211's optical and X-ray light curves which showed it to be an ADC -- to observe X-ray bursts with photospheric expansion the neutron star surface has to be visible, but in ADC sources the neutron star is hidden from view by the accretion disc. This problem goes away if it is the second LMXB, and not AC211, which is the burster. The {\it Chandra} discovery may solve one mystery but it produces another: the fact that AC211's X-ray luminosity is even fainter than previously believed makes its unusually high optical luminosity even more puzzling, and suggests that the central X-ray source hidden from view must be exceptionally luminous, possibly indicating a very high mass-transfer rate from the companion star.

Determining whether AC211 has an extremely high mass-transfer rate and is in an unusual evolutionary state is important: it is relevant to the understanding of LMXB evolution and to the understanding of stellar interactions within, and the evolution of, globular clusters. Theoretical determinations of the numbers of neutron stars in globular clusters and the efficiency with which they interact with stars in the cluster cores to form binaries, combined with the large number of millisecond radio pulsars (end-products of LMXB evolution) observed in the clusters, imply that we see far fewer LMXBs in globular clusters than we should. This may imply that the evolution of LMXBs within globular clusters is accelerated, and that their lifetimes are very short (less than 10$^8$ years; Hut, Murphy \& Verbunt 1991), which in turn implies high mass-transfer rates.

Not much is known about either the compact object or the donor star comprising AC211, although its location in the core of a globular cluster enables us to make reasonable assumptions about their nature. No globular cluster X-ray binary is known to contain a black hole, and neutron stars are abundant in globular clusters: theory predicts that the number of neutron stars in a dense globular cluster such as M15 is \til 4000 (Hut, Murphy \& Verbunt 1991). The chances therefore that the compact object in AC211 is a neutron star are high. The system has an orbital period of 17.1 hours, indicating that it is unlikely that the companion star is on the main sequence (at that orbital period a main sequence donor star would not fill its Roche lobe). The main sequence turn off point for M15 occurs at \til 0.8 M$_{\odot}$ (Fahlman et al. 1985), so we assume that AC211's donor star is an 0.8-M$_{\odot}$ star which has begun to evolve off the main sequence.

\section{Previous observations and models}

\subsection{The Absorption Line Profiles: Structured Mass Outflow?}

The most striking features in the spectra of AC211 are the blue-shifted absorption components seen in the Balmer and \he lines. The line centres are blue-shifted with velocities ranging from 150 to 800 km~s$^{-1}$ with respect to the cluster core (Naylor et al. 1988, Ilovaisky 1989). Initially these high blue-shifts led Naylor et al. (1988) to propose that AC211 is the product of a 3-body interaction, and that its formation resulted in its ejection from the cluster core.

However, Ilovaisky (1989) showed that the \he and \hb absorption lines were complex in structure, each consisting of a blend of two or three Gaussian absorption components of different strengths and velocities. The least blue-shifted absorption component in the \he line, which we assume is formed somewhere on or close to the components of the binary system (possibly on the accretion disc rim), had a velocity close to the cluster velocity, which means that AC211 is not being ejected from the cluster core. The multiple-component nature of AC211's absorption lines indicates that we are seeing some kind of structured outflow from the system. 

Outflow from the binary system has been a popular idea ever since Charles, Jones \& Naylor (1986) determined a value for the mass transfer rate of $\sim 10^{-8} {\rm M}_{\odot}~{\rm y}^{-1}$, which is extremely high. A high mass transfer rate for AC211 is also indicated by its large orbital period, which requires an evolved donor star. Evolved stars tend to give higher mass transfer rates: evolutionary calculations for mass transfer from the donor star, derived by Rappaport, Verbunt \& Joss (1983), give mass transfer rates of
\begin{equation}
\dot{M} = 2 \times 10^{-11} P_{\rm orb}^{3.2}(h) {\rm M}_{\odot}~{\rm y}^{-1}.
\end{equation}
An alternate derivation by McDermott \& Taam (1989) gives, 
\begin{equation}
\dot{M} = 2 \times 10^{-11} P_{\rm orb}^{3.7}(h) {\rm M}_{\odot}~{\rm y}^{-1}.
\end{equation}
These equations give $\dot{M}$ values for AC211 of $\sim 1.8 \times 10^{-7} {\rm M}_{\odot}~{\rm y}^{-1}$ and $\sim 7.3 \times 10^{-7} {\rm M}_{\odot}~{\rm y}^{-1}$ respectively, both of which substantially exceed the Eddington-limited value of $\sim 1.6 \times 10^{-10} {\rm M}_{\odot}~{\rm y}^{-1}$ for a $1.4{\rm M}_{\odot}$ neutron star (Frank, King \& Raine 1992). This indicates that most of the material flowing from the companion star never accretes onto the compact object, but must somehow escape the binary system; hence the interest in mass-outflow mechanisms in AC211.

Two mass-outflow models have been proposed to attempt to explain the velocities of AC211's absorption lines: an accretion disc wind (Fabian, Guilbert \& Callanan 1987) and a common envelope (Bailyn, Garcia \& Grindlay 1989), discussed in detail below.

\subsection{The Accretion Disc Wind Hypothesis}

Based on X-ray observations of AC211, Fabian, Guilbert \& Callanan (1987) proposed an ionized accretion disc corona with a Thomson scattering depth $\tau_{\rm T} > 1$ extending to R $\sim 10^{11}$ cm from the compact object (the radius of the compact object's Roche-lobe is $\sim 10^{11}$ cm). Beyond $\sim 10^{11}$ cm, they argue, the corona cannot be maintained by X-ray heating, and the coronal gas must be in the form of a cold ($T \leq 10^5$~K), supersonic wind with velocities $\geq$ 300 km s$^{-1}$. They calculate the wind mass-loss rate to be in excess of $10^{-7} {\rm M}_{\odot}~{\rm y}^{-1}$.

Naylor et al. (1988) argue that the \he and \heii line profiles are inconsistent with the Fabian, Guilbert \& Callanan model, and that the required mass-loss rate in the wind of $\sim 10^{-7}{\rm M}_{\odot}~{\rm y}^{-1}$ is implausibly high. However, Naylor et al. base their argument on the assumption that the \he absorption and \heii emission are formed in the same gas. This is not necessarily true: the \he and \heii line profiles are extremely complex, and it is possible that some components of the lines may be formed in an ADC wind beyond $\sim 10^{11}$ cm as Fabian, Guilbert \& Callanan suggest, while other line-forming regions exist elsewhere in the system, for example the accretion disc. We will return to the disc wind scenario in \S 4.

\subsection{The Common-Envelope/L$_2$-Outflow Hypothesis}

\begin{figure}
\hspace{0.4cm}
\psfig{file=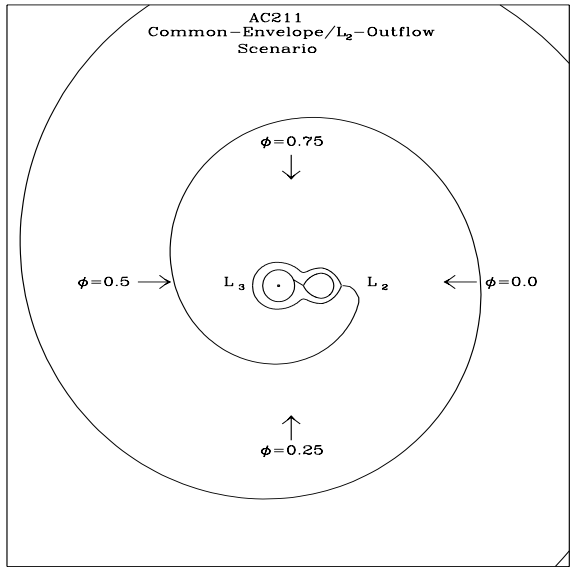,width=7.5cm,height=7.5cm,clip=}
\caption{Schematic of Bailyn, Garcia \& Grindlay's (1989) common envelope model for AC211 in M15. The
geometry of the outflow from the L$_2$ point as the binary system rotates anti-clockwise gives rise to a `garden-sprinkler' effect, in which a spiral of gas moves outwards in the orbital plane. Adapted from BGG.}\label{fig:spiral}
\end{figure}
\begin{figure} 
\psfig{file=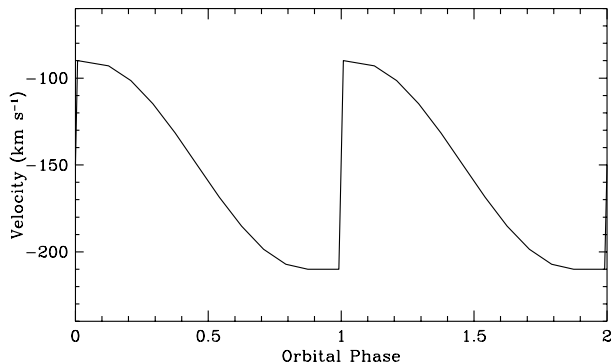,width=8cm,clip=}
\caption{Schematic diagram showing the radial-velocity behaviour of absorption lines originating in the outflowing spiral of material flowing from L$_2$. Immediately before mid-eclipse (phase 0.0) the absorption is dominated by material which, in the rotating frame, has already circled the binary once and has a higher velocity than material leaving L$_2$. Maximum blue shift therefore occurs just before phase 0.0. Adapted from BGG.}\label{fig:rvmod}
\end{figure}

The explanation that has become the favoured model over the last decade is that of Bailyn, Garcia \& Grindlay (1989) (hereafter BGG) who propose a common-envelope scenario in which the blue-shifted lines are formed in material flowing through the second Lagrangian point (L$_2$), located at the back of the companion star (Fig.{\th}\ref{fig:spiral}). On evolutionary grounds, BGG argue that AC211 consists of a 1.4 ${\rm M}_{\odot}$ neutron star and an 0.8 ${\rm M}_{\odot}$ secondary which has recently left the main sequence (the main sequence turn-off point for M15 is at $\sim 0.8 {\rm M}_{\odot}$; Fahlman et al. 1985). They suggest that tidal capture created a system close enough that mass transfer began very soon after the companion star left the main sequence. BGG argue that at this mass transfer is initially thermally unstable, resulting in the accumulation of material between the companion's Roche lobe and the L$_2$ surface. As soon as sufficient material has transferred from the companion to the primary to move the mass ratio $M_2/M_1$ out of the unstable regime, BGG suggest that stable, super-Eddington mass transfer sets in which maintains the common envelope.

Fig.{\th}\ref{fig:spiral} is a schematic of the common-envelope/L$_2$-outflow model proposed by BGG and the resulting mass-outflow geometry. Material flowing over the L$_2$ saddle-point and escaping the system would spiral steadily outwards (the so-called `garden-sprinkler' effect). At each phase in the binary orbit an observer would therefore see a blue-shift in absorption lines originating within the outward-spiraling material. BGG model the radial velocity curve produced by absorption lines originating within this outflowing material; we show a schematic diagram of their modelled radial-velocity curve in Fig.{\th}\ref{fig:rvmod}. Note that for all phases there is a blue-shift, and that, for an L$_2$-outflow scenario, the maximum blue-shift has to occur just before phase 0.0, when the absorption is dominated by material which, in the rotating frame, has already circled the binary once and has a higher velocity than material leaving L$_2$.

In modelling the radial velocity curve represented in Fig.{\th}\ref{fig:rvmod}, BGG assume that the material flowing from L$_2$ has a temperature of 10,000 K (close to the recombination temperature of helium). This low temperature is not unreasonable: the material at L$_2$ is shielded from irradiation from the compact object and hot inner disc by the companion star. Furthermore, the outflowing spiral, because it remains in the orbital plane, is shielded by the rim of the accretion disc. BGG find that the spreading of the particles is small enough to preserve the identity of the first few turns of the outward-moving spiral, but find that gas from outside the innermost turn of the spiral contributes only marginally to the integrated \he absorption.

BGG derive a plausible estimate for the mass-outflow rate $\dot{M}_{\rm spiral}$ of the spiral which gives a column of neutral He sufficient to produce the observed \he absorption. Using an order-of-magnitude Str\"omgren sphere-like calculation, they derive the following expression for $I_{\rm He}$, the number of ionizations of He per second, in the `wedge' of gas formed by the spiral expanding in the equatorial plane:
\begin{equation}
I_{\rm He} = \alpha \int_{R_0}^{R_s} N^2(r)4\pi r^2 dr = \frac{\alpha \dot{M}^2_{\rm spiral}}{v^2} ({R_0}^{-1} - {R_s}^{-1}),
\end{equation}
where $\alpha N^2$ is the number of recombinations of He per cm$^3$, $\alpha$ is the recombination cross-section for states other than the ground state; $\alpha \sim 2\times 10^{-13}$ for He at 10,000 K (Osterbrock 1974), $R_0$ = $R_{{\rm L}_2} \sim 1.45 {\rm R}_{\odot} \sim 10^{11}$ cm, $R_s$ is the radius of the Str\"omgren ``wedge'' inside of which He will be ionized, and $v=150~{\rm km ~s}^{-1}$, the mean observed velocity of the \he line. They assume that 1/4$\pi$ of the observed luminosity passes through the outflow spiral (which is shielded from AC211's intrinsic luminosity by the accretion disc), and that $\sim$ 1/10 of these photons interact with a helium rather than a hydrogen atom (ionization is assumed to be efficient even though most photons are in the form of X-rays; the excess energy will result in kinetic motion, leading to collisional ionizations). Assuming $R_s$/$R_0$ = 1.1 (to ensure an insignificant column of ionized material), they find that $\dot{M}_{\rm spiral} \sim 10^{-9} {\rm M}_{\odot}~{\rm y}^{-1}$.

Although this common envelope scenario has some very appealing features, and has for a long time been a very
popular explanation for AC211's peculiar behaviour, it has some potentially serious
drawbacks. It is
incompatible with the observed extensive corona: it is not clear how
cooler, denser common envelope material can lie above a very hot,
low-density corona and remain stable, unless most of the common-envelope is
somehow confined to the orbital plane.

\section{Reanalysis of Archival Data}

\begin{figure} 
\psfig{file=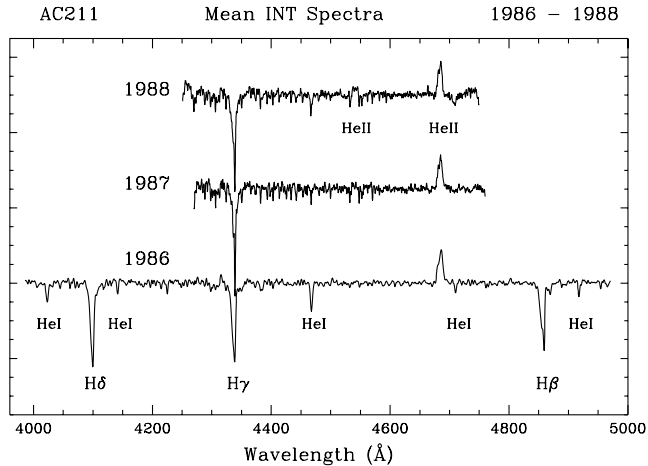,width=8.5cm,clip=}
\caption{Mean spectra of AC211 obtained with the INT from 1986 to 1988.}\label{fig:3specs}
\end{figure}

In all the early work published on AC211 (e.g. Naylor et al. 1988; BGG; Ilovaisky 1989), the system's orbital period was believed to be $\sim$ 8.5 hours. Extensive observations by Ilovaisky et al. (1993), and eclipse analyses of archival X-ray data by Homer \& Charles (1998), have since revealed the true orbital period to be 17.11 hours. Using the most up-to-date ephemeris for AC211 (Ioannou et al. 2001), we have re-examined the original spectroscopy obtained by PAC and TN in three sets of observing runs: August 1986 (published
in Naylor et al. 1988), August 1987 (published in Naylor \& Charles
1989), and August 1988 (unpublished). A summary of the observations is given in Table 1. The data were obtained with the RGO Intermediate Dispersion Spectrograph and IPCS (1986 \& 1987) or CCD (1988) on the Isaac Newton Telescope at the Roque de los Muchachos Observatory on La Palma.
Each dataset consists of three or four nights of 600-s exposures. The mean spectrum for each year is plotted in Fig.{\th}\ref{fig:3specs}. A detailed description of how the spectra were extracted from the IPCS data can be found in Naylor et al. (1988).

\begin{figure} 
\psfig{file=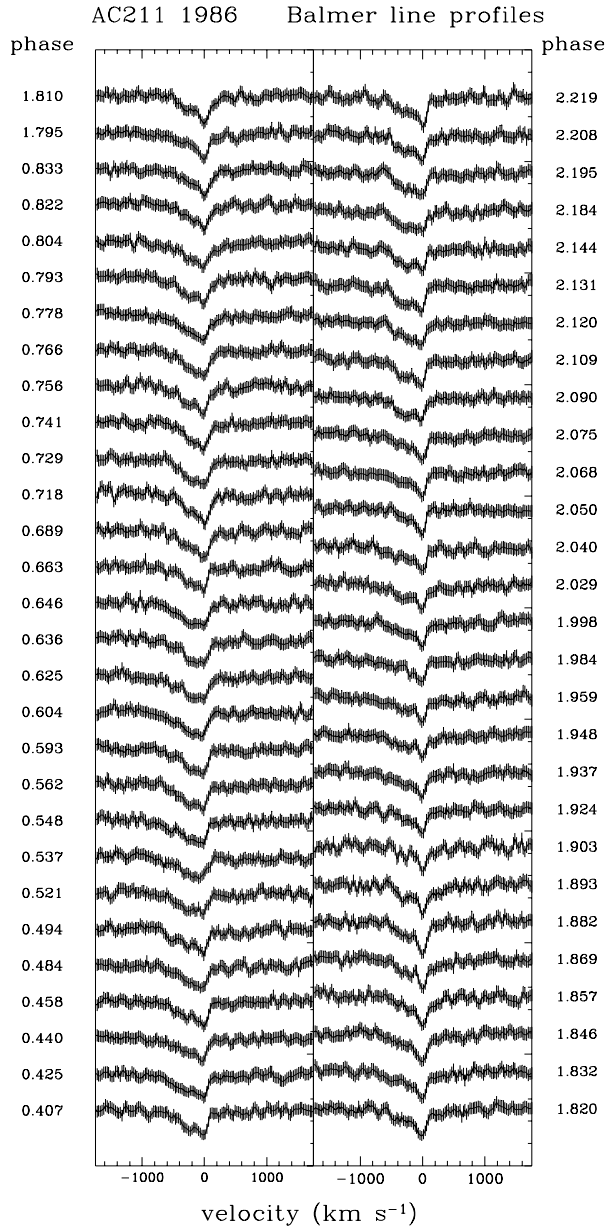,width=8.0cm,height=16.7cm,clip=}
\caption{Phase dependence of AC211's Balmer lines in the 1986 spectra. Phases are given in orbital cycles from HJD 2446645.140196. For display purposes only (to reduce errors and to highlight trends) each profile is the velocity-binned sum of the H${\beta}$, H${\gamma}$ and H${\delta}$ lines in each spectrum. Individual H${\beta}$, H${\gamma}$ and H${\delta}$ profiles differ slightly within each spectrum (for example, emission infilling of the red wing becomes more prominent from H${\delta}$ to H${\beta}$), but are sufficiently similar that summing them is useful for highlighting orbital-phase-dependent trends. $v=0$ km~s$^{-1}$ corresponds to the cluster velocity; the absorption component at $v=0$ km~s$^{-1}$ in each profile is the cluster background.}\label{fig:stack}
\end{figure}

\begin{table} 
\begin{center}
\caption{Summary of Observations}

\begin{tabular}{cccc}
\hline
Date             & Spectral              & Dispersion & Number of \\
                 & Range                 & (\AA~pix$^{-1}$) & Spectra   \\
\hline					              
      01/8/1986  & 3250 \AA ~-- 5175 \AA & 1.00  & 15 \\
  02--03/8/1986  & 3950 \AA ~-- 4975 \AA & 0.50  & 58 \\
  15--18/8/1987  & 4270 \AA ~-- 4770 \AA & 0.25  & 79 \\
  04--06/8/1988  & 4270 \AA ~-- 4770 \AA & 0.25  & 98 \\
\hline
 
\end{tabular} 
\end{center}
\end{table} 

\subsection{The Balmer Lines}

AC211, located in the core of M15, is situated in an exceptionally crowded field, which makes observing it a considerable technical challenge, and requires superb seeing. Ground-based spectra of AC211 are heavily contaminated with light from the cluster core. Variations in seeing between and during exposures leads to changes in the relative contribution of the cluster background in AC211's spectra, which makes it impossible to determine AC211's continuum flux and hence obtain reliable equivalent widths for its absorption and emission lines.

\begin{figure} 
\psfig{file=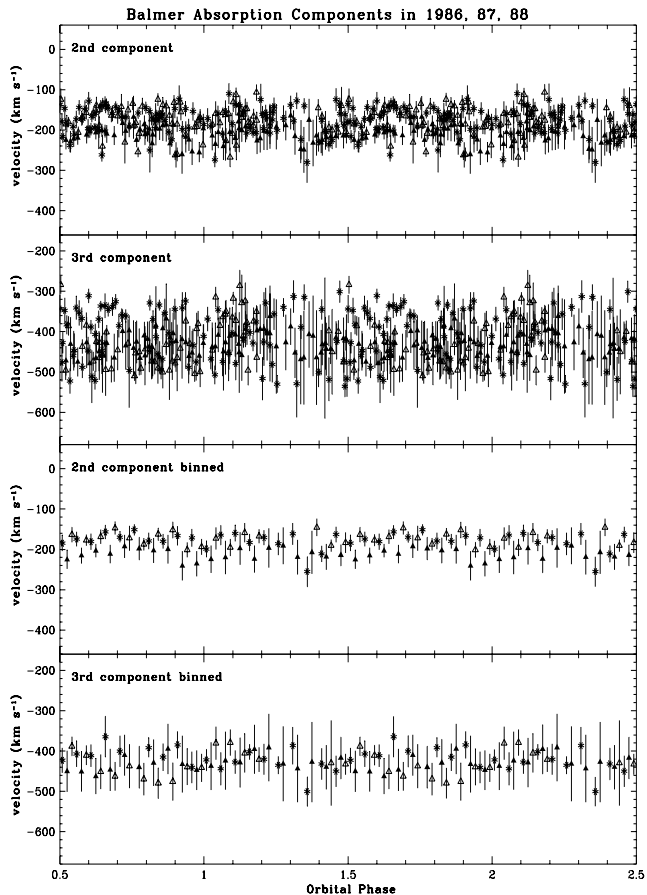,width=8.5cm,height=12cm,clip=}
\caption{Radial velocities of the Balmer lines, folded on
the correct (17.1-h) orbital period, from archival spectra of AC211 in M15. The data are plotted twice for clarity. The Balmer lines are complex, consisting of three discrete absorption components. Component 1 (not plotted) is always at rest with respect to the cluster, and is most likely almost entirely due to contamination from cluster stars. There is evidence for infilling of the red wings, indicating the presence of P Cyg emission components from AC211. Components 2 and 3 are blue shifted $\sim$ 200 km~s$^{-1}$ and $\sim$ 420 km~s$^{-1}$ respectively. Filled triangles are 1986 data, open triangles are 1987 data, and stars are 1988 data.}\label{fig:balmer}
\end{figure}

The cluster contamination is clearly visible in the Balmer lines. In Fig.{\th}\ref{fig:stack} we have for clarity merged the H${\beta}$, H${\gamma}$ and H${\delta}$ lines, to reduce visual errors in order to highlight orbital-phase-dependent changes in the Balmer line profiles. As noted by Ilovaisky (1989), the Balmer line profiles are complex in structure. The strongest feature in the line profiles is the relatively narrow (FWHM \til 3 \AA) absorption component (at $v=0$ km~s$^{-1}$ in Fig.{\th}\ref{fig:stack}) which corresponds to the cluster background contamination. Other prominent features in the Balmer profiles are the broad blue absorption features and the infilling due to emission in the red wing of the cluster background absorption lines. Ilovaisky (1989) finds strong red-wing emission in the Balmer lines: his data were obtained on the 3.6-m Canada-France-Hawaii telescope in excellent seeing conditions; therefore the cluster background contamination in his spectra is much smaller than in our INT data, where the red-wing emission in the Balmer lines is swamped by the cluster light.

We analysed all the Balmer lines in each spectrum using a multi-Gaussian routine, based on Marquardt's method of minimization, within the {\sc molly} spectral analysis package of T. R. Marsh. We fitted various combinations of Gaussians to the line profiles to try to model simultaneously the cluster absorption, the red wing emission infilling and the broad blue absorption wing. We found that the models with the best ${\chi}^2$ consisted of an absorption component at the cluster velocity and two additional blue-shifted absorption components. We find absorption component 2 to have an average blue shift of \til 200 km~s$^{-1}$, and \til 440 km~s$^{-1}$ for absorption component 3. Our results are similar to those of Ilovaisky, who finds (with data of considerably higher signal-to-noise) that the blue wings of the Balmer lines consist of two or more distinct absorption components with velocities ranging from 150 to 800 km~s$^{-1}$ blue-wards of the cluster background absorption component. The radial velocities of the two blue components are plotted in Fig.{\th}\ref{fig:balmer}. What is immediately apparent is that there is no obvious orbital phase dependency.

The signal-to-noise of our data is too poor to allow us to investigate the behaviour of the widths of the blue absorption components: in our multi-Gaussian analysis we fixed the FWHM of the components using mean values from the Ilovaisky data.

The Balmer radial velocities present two challenges: it is difficult to see how the lines can be free of any phase dependence, and it is difficult to imagine what could give rise to the two distinct blue-shifted absorption components (if real). It is possible that the Balmer absorption profiles represent something more complex than our simple multi-Gaussian models allow for. With so many possible free parameters and the low signal-to-noise of our line profiles, degeneracy cannot be excluded and our minimum-${\chi}^2$ fits may not represent the intrinsic nature of the line profiles. If our radial velocity results are real, then we are dealing with some kind of highly structured mass outflow: the velocities appear to indicate expanding shells or rings. The physical mechanism that might produce such structures from an X-ray binary is a mystery.

Intriguingly, Torres et al. (2001) have found, in MMT spectra of AC211 obtained in 1996, that the Balmer line velocities do show some evidence of phase dependency in 1996, and that their greatest blue shift is at phase $\sim 0.25$. Other studies (van Zyl et al. in preparation) find that the blue-shifted absorption components of the \ha line profiles in HST/STIS observations of AC211 show no orbital variation in velocity at all. It appears that AC211 can exhibit different states, in which the Balmer line profiles sometimes have an orbital-phase dependence and at other times not. P-Cygni-profiled Balmer lines that show no correlation with orbital phase have been observed in other interacting binaries, for example in the cataclysmic variable BZ Cam (Ringwald \& Naylor 1988), but the reasons for this behaviour are unknown.

\subsection{The \he Line}

\begin{figure} 
\psfig{file=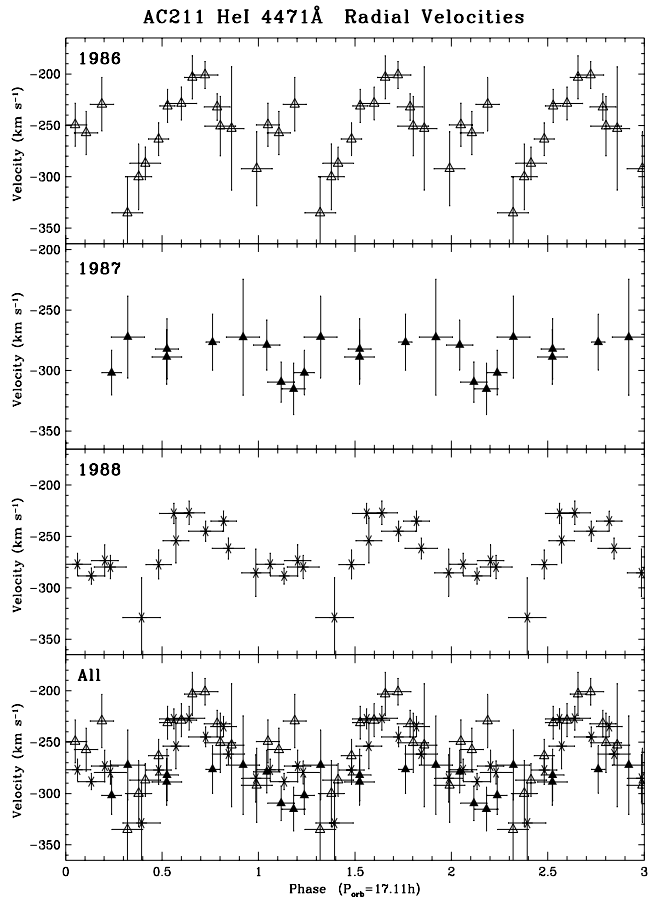,width=8.5cm,height=12cm,clip=}
\caption{Revised radial velocities of the HeI 4471 \AA~ line, folded on
the correct 17.1-h orbital period. The data are plotted 3 times for clarity. The cluster velocity of -115 km~s$^{-1}$ has not been removed. The mean velocity and semi-amplitude are in remarkable agreement with the velocities predicted by BGG, but the maximum blue shift does not coincide with phase 0.0, as required by a common envelope/L$_2$-outflow model.}\label{fig:heI}
\end{figure}

The signal-to-noise in the \he line is poor, especially in the 1987 and 1988 data with their higher resolution. We have therefore summed the spectra into (roughly) hour-long means. These hour-long mean spectra do not unfortunately have a signal-to-noise high enough to allow multi-Gaussian fitting of the \he line; the two discrete absorption components reported by Ilovaisky are not resolved in our data. We have had to determine the radial velocities by fitting only a single Gaussian to the \he line. This should nonetheless still give us information about the radial velocity behaviour of the line: Ilovaisky's work shows that one absorption component remains nearly constant. This indicates that our single-Gaussian velocities will give us the shape of the \he radial-velocity curve, but that the values we obtain for the mean velocity and semi-amplitude will be too low. 

Ilovaisky also reports P Cygni behaviour in the \he line. We do not detect any evidence of a red emission component in our data, probably because of our poorer signal-to-noise (in Ilovaisky's data the emission component at its strongest is less than 1/4 the strength of the absorption components). The possible presence of P Cygni features further complicates any radial velocity study of the \he line.

In Fig.{\th}\ref{fig:heI} we plot AC211's \he-line radial velocities, folded, for the first time, on the correct (17.1-h) orbital phase. All previous radial velocity studies of AC211 have assumed the incorrect 8.5-h period. We make use of the most recent ephemeris derived for AC211 (Ioannou et al. (2001), an extension of work by Ilovaisky et al. (1993) and Homer \& Charles (1998). Three orbital cycles of the folded data are plotted for clarity.

Some striking features of the \he-line radial velocities are immediately evident. First, the velocities display a strong orbital-phase dependency, completely unlike the Balmer absorption lines. Secondly, it is also apparent that there is a substantial long-term temporal variation in the shape of the radial-velocity curve. The most significant feature however, from a mass-outflow point of view, is that the phase of maximum blue-shift does not occur at phase zero (i.e., when the compact object and disc are behind the companion star), as required by BGG for L$_2$-outflow, but rather near phase 0.25. This presents a serious challenge to the plausibility of the common envelope model. A discussion of the possible sites of \he formation is given in \S4.

\begin{figure} 
\psfig{file=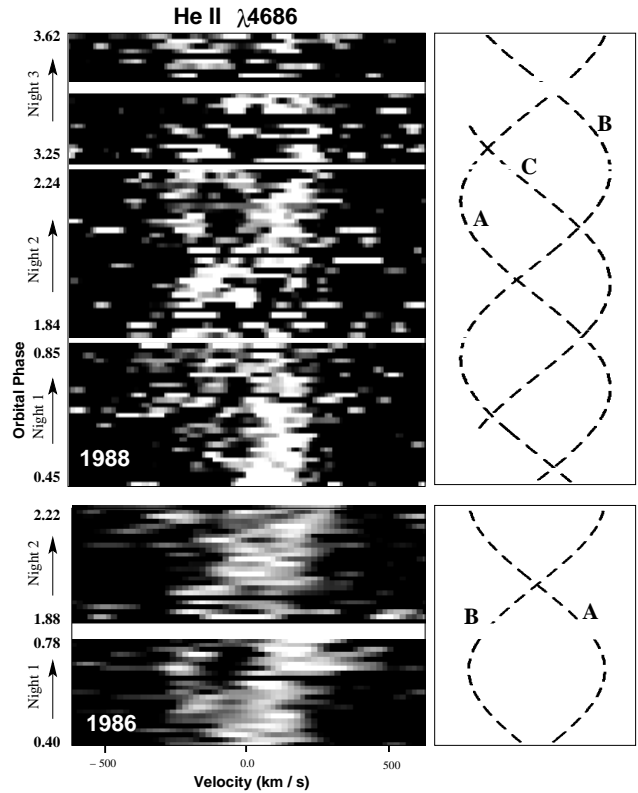,width=8.3cm}
\caption{Trailed spectrograms of the \heii line. Plotted are two nights from 1986 and three nights from 1988. The orbital phase coverage of each night is indicated on the left. On the right we plot a sketch of the emission components that appear to be present. Components A and B are consistent with the motion of the compact object and an irradiated companion star respectively. }\label{fig:heII}
\end{figure}

\subsection{The \heii Line}

Fig.{\th}\ref{fig:heII} is a trailed spectrogram of AC211's \heii line, taken from two nights in 1986 and three in 1988; the nights on which the signal-to-noise is greatest. The line clearly exhibits complex behaviour. 

\begin{figure*}
{\hspace{0.1cm}\psfig{file=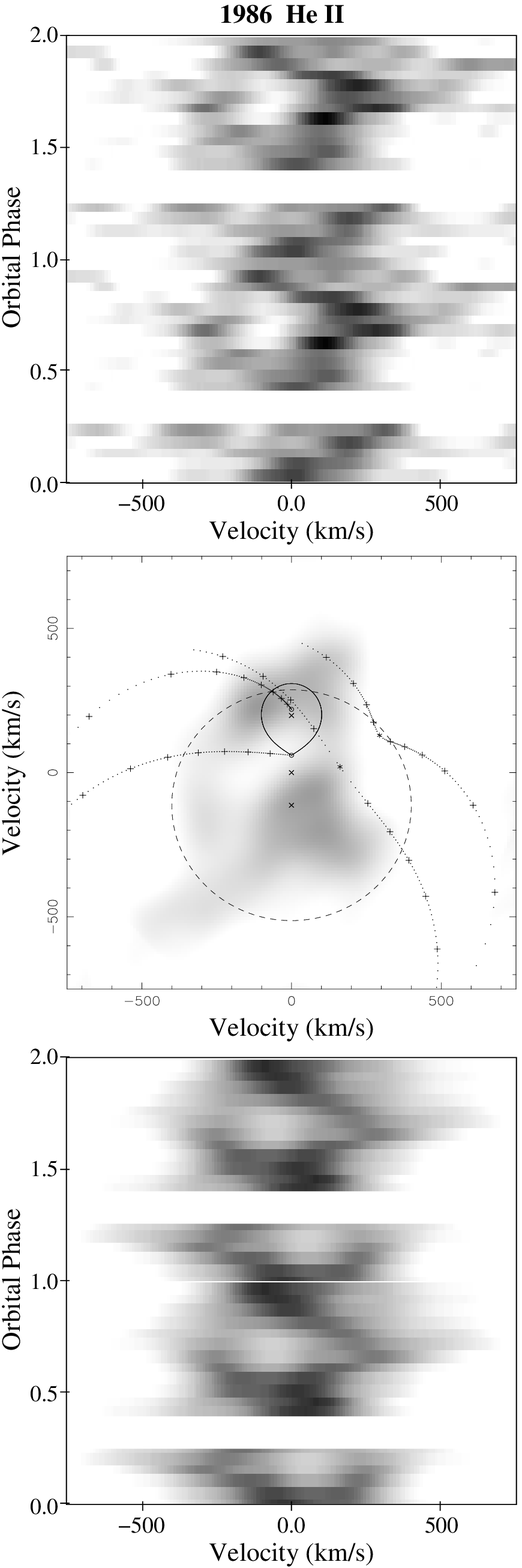,height=18cm} 
\hspace{2cm} 
\psfig{file=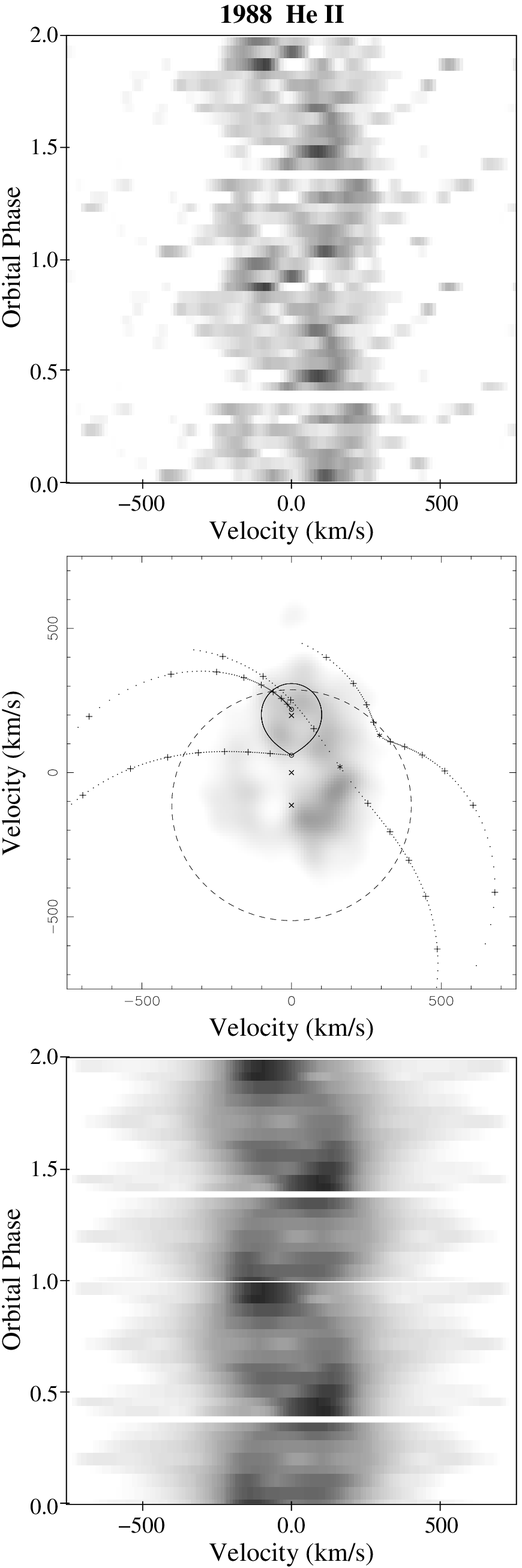,height=18cm}}\hspace{0.2cm}
\caption{Doppler tomograms of the \heii lines in 1986 and 1988. The top panels show spectrograms of the orbital-phase-binned data, and the bottom panels show the spectrograms reconstructed from the tomograms. Phase increases from bottom to top, and each spectrogram is plotted twice for clarity. The Doppler maps show the Roche lobe of the companion, with crosses indicating the centres of mass of the companion, neutron star and system. A mass ratio of $M_2/M_1 = 0.8~{\rm M}_{\odot} ~/~ 1.4~{\rm M}_{\odot} \sim 0.57$ is assumed. The ring indicates the position in velocity space of the outer accretion disc ($\sim 0.85 R_{\rm L}$). A full orbit of the accretion stream is plotted.}\label{fig:doppler}
\end{figure*}

The 1988 spectrogram appears to show three distinct sinusoidal components, which we have labeled A, B and C in Fig.{\th}\ref{fig:heII}. Component B has maximum blue and red shifts near phase $\sim$ 0.75 and $\sim$ 0.25 respectively, and a semi-amplitude of $\sim$ 200 km~s$^{-1}$. The blue part of component B is very weak on night 1, and seems to disappear between phases $\sim$ 0.45 and $\sim$ 0.65, but on night 3 it is strong at these phases. AC211 evidently exhibits substantial variations in behaviour from orbit to orbit.

The phasing of component B, which is also clearly present in the 1986 spectrogram, is consistent with the motion of an irradiated companion. Assuming the canonical value of 1.4 ${\rm M}_{\odot}$ for the mass of a neutron star primary and 0.8 ${\rm M}_{\odot}$ for the mass of the companion (corresponding to the main sequence turn-off point for M15), and assuming an inclination of 90$^{\circ}$ for the system, Kepler's laws give a velocity for the companion of 198 km~s$^{-1}$, which coincides very closely with the amplitude of component B. If we are indeed seeing recombination \heii from the X-ray-irradiated face of the companion star, it would be the first time that the companion of AC211 has been detected. 

Component A is substantially broader and stronger than B, and its phasing and velocity correspond to the motion of the compact object. This component could come from the hot inner regions of the accretion disc, from a disc wind, or from the accretion disc corona, which will all have the same phase-dependence as the compact object. This component may be formed predominantly in AC211's large accretion disc corona or disc wind, because the \heii emission is never eclipsed, indicating that the \heii-emitting region is substantially larger than the companion star. Indeed, work by Ioannou et al. (2001) on HST UV spectra, and Torres et al. (2001) on optical spectra, shows that the equivalent width of the \heii lines increases sharply during eclipse as the continuum flux drops, supporting the idea that the \heii emission is formed in a large corona. We expect the accretion disc corona of AC211 to be particularly large because of M15's (and therefore AC211's) very low metallicity (Fabian, Guilbert \& Callanan 1987).

Component C, a third, weaker component in the 1988 spectrogram, appears most strongly between phases 0.1 and 0.3. Its phasing suggests that it may come from an accretion stream overflowing the disc. A similar \heii emission component, attributed to an overflowing stream, has been observed in the SW Sex star V1315 Aql (Hellier 1996). The velocity amplitude of component C is much lower than what one would expect from a ballistic stream; however at the phases at which it is most clearly observed, an overflowing stream would be viewed almost perpendicularly, and therefore its line-of-sight velocity would be low. In addition, Armitage and Livio (1998) show that an overflowing stream would have a velocity well below free-fall.

\subsection{Doppler Tomograms}

Fig.{\th}\ref{fig:doppler} shows the trailed spectra and Doppler tomograms of the \heii line. The signal-to-noise in the lines is poor, but the multiple-component structure of the line profiles is evident in the 1986 and 1988 data (the 1987 dataset is the victim of poor weather; we have not included it in Fig.{\th}\ref{fig:doppler}). Unlike in Fig.{\th}\ref{fig:heII}, the spectra have been phase-binned. The Roche lobe of the companion and the centres of mass of the companion, neutron star and binary system are indicated; we have assumed a mass ratio of $M_2/M_1 = 0.8~{\rm M}_{\odot} ~/~ 1.4~{\rm M}_{\odot} \sim 0.57$. We have plotted a full orbit of the accretion stream, and the ring indicates the position in velocity space of the outer accretion disc, assumed to be $\sim 0.85 R_{\rm L}$.

In these \heii tomograms there is no evidence of emission from the accretion disc or bright spot: the characteristic ring and the bright-spot signature in the upper-left quadrant typically seen in tomograms of interacting binaries is entirely absent here. In this respect these tomograms resemble those of the SW Sex class of CVs. SW Sex tomograms, however, tend to be bright in the lower-left quadrant (believed to be the signature of an overflowing accretion stream re-impacting the disc), while these data are bright in the lower-right quadrant, a position in velocity-space which does not coincide with any obvious sites of excess emission within the binary's geometry. The very low velocity of this emission indicates that it may be formed in a bipolar disc wind viewed almost edge-on, or in the ADC, although the off-set to the right (if real) is puzzling.

Our tomograms also show emission in regions associated with the mass donor, which is consistent with the suggestion in the previous subsection that we may have detected the irradiated companion. Component C in Fig.{\th}\ref{fig:heII}, possibly a signature of disc overflow, appears as a faint feature to the left of the compact object in the 1988 tomogram, in a region associated with disc overflow in SW Sex stars. Intriguingly, there also appears to be a brightening in this area in the 1986 tomogram.

\section{Discussion}

BGG's common envelope model implies that outflow from L$_2$ would result in the HeI lines having a maximum blue shift at phase 0.0. A maximum blue-shift at phase $\sim 0.25$, which our reanalysis of the original radial velocity study has revealed, would appear to be incompatible with this model. However, BGG's modelling involves many simplifications and approximations, and a more extensive development of this option may yield somewhat different results.

If our new results do rule out an L$_2$-outflow scenario, an alternative explanation for the blue-shifted absorption lines will have to be found. We consider four possible alternative scenarios:

\begin{itemize}
\item
Material flowing through the L$_3$ point (located on the opposite side of the binary system to L$_2$) would have travelled through the compact object's Roche lobe on a ballistic trajectory, and therefore {\it would} show a maximum blue-shift near phase $\sim 0.25$. Intense irradiation of an overflowing accretion stream by the compact object and inner disc may give stream material enough energy to escape the system through the L$_3$ saddle-point.

\item
An L$_3$-outflow may be the result of a magnetic propeller, whereby overflowing stream material is accelerated out of the system by the magnetic field of a rapidly-spinning compact object (e.g. Wynn, King \& Horne 1997), or by the magnetic field of the inner accretion disc (Horne 1999). Hynes et al. (2001) claim to have observed such a magnetic propeller in action in the LMXB XTE~J2123-058.

\item
Groot et al. (2001) show that the lines formed at the stream-disc impact site have maximum blue-shifts at phases 0.25 -- 0.4, consistent with our \he results. A sufficiently energetic stream-disc impact site (expected for AC211's inferred high mass transfer rate) could produce the temperatures and material necessary for \he absorption. However, for the line to be visible in eclipse, the impact region would require a vertical extent comparable to that of the companion star. Intriguingly, the 1986 (and perhaps also the 1988) \he velocities in Fig.{\th}\ref{fig:heI} show a sharp discontinuity at phase 0.1, which is when the impact region re-emerges.

\item 
Material flowing out from an accretion disc wind would be expected to have a maximum blue-shift which coincided with the compact object's approach velocity, i.e. at phase $\sim 0.25$. HST/STIS UV observations of \civ (Ioannou et al. 2001) show that AC211 does have an accretion disc wind. However, the wind velocity (\til1500~km~s$^{-1}$) shows no detectable variation with orbital phase.

\end{itemize}
Apart from the final item, the above scenarios (along with BGG's L$_2$-outflow scenario) still require extensive and rigorous modelling before they will have sufficient predictive power to be of use. Accretion disc winds, however, have been modelled extensively for over two decades (e.g. Drew 1987; Begelman, McKee \& Shields 1983). Nevertheless, it is difficult to understand AC211's \he absorption in terms of an accretion disc wind, mainly because X-rays should over-ionize the wind (Drew 2002, private communication).

\section{Conclusions}

The evolutionary state of X2127+119/AC211 is of great interest because of theoretical predictions that globular cluster LMXBs evolve much faster than their galactic cousins (Hut, Murphy \& Verbunt 1991). That AC211 appeared to be embedded in a common envelope seemed to imply that at least one globular cluster LMXB did have an unusual evolutionary history. Our results, however, show that AC211 is unlikely to be a common envelope system because the \he radial velocities appear to be inconsistent with an L$_2$ outflow. We may also have detected the irradiated face of the elusive companion star in the \heii line profiles. 

The Balmer and \he lines indicate the presence of outflows from the system, but without better data it is difficult to determine their exact nature. There are several possible scenarios which may explain the nature of these outflows, but each requires rigorous modelling before it can be considered seriously as a mechanism for producing our observed \he line profiles.

\section*{Acknowledgments}

This paper is based on observations made with the Isaac Newton Telescope operated on the island of La Palma by the Isaac Newton Group in the Spanish Observatorio del Roque de los Muchachos of the Instituto de Astrofisica de Canarias.

LvZ acknowledges the hospitality of the University of Southampton, where much of this paper was written. LvZ would like to thank Rob Hynes, Danny Steeghs, Paul Groot and Coel Hellier for useful discussions on the nature of AC211 and on Doppler tomography, and Tom Marsh for the use of his {\sc molly, doppler} and {\sc trailer} spectral analysis software. LvZ acknowledges the support of scholarships from the Vatican Observatory, the National Research Foundation (South Africa), the University of Cape Town, and the Overseas Research Studentship scheme (UK).

TN acknowledges the support of a PPARC Advanced Fellowship during part of this work.

We would also like to thank the anonymous referee for helpful comments.

\label{lastpage}

\end{document}